\documentclass{elsart}

\usepackage{harvard}

\usepackage{graphicx}

\usepackage{amssymb}

\def\url#1{{\ttfamily\def\/{/\discretionary{}{}{}}#1}}

\begin{document}

\begin{frontmatter}
\title{Simulations of Nonthermal Electron Transport in Multidimensional Flows: Synthetic Observations of Radio Galaxies}


\author[Minn]{I.L. Tregillis\thanksref{it}}, 
\author[Minn]{T.W. Jones\thanksref{twj}},
\author[Ryu]{Dongsu Ryu\thanksref{ryu}},
\author[Park]{Charles Park\thanksref{cpark}}

\thanks[it]{E-mail: tregilli@msi.umn.edu}
\thanks[twj]{E-mail: twj@astro.spa.umn.edu}
\thanks[ryu]{E-mail: ryu@canopus.chungnam.ac.kr}
\thanks[cpark]{E-mail: charles.park@yale.edu; Department of Mathematics,
Yale University, 10 Hillhouse Avenue, P.O. Box 208283, New Haven, CT 06520-8283}

\address[Minn]{School of Physics and Astronomy, University of Minnesota, 
116 Church St. S.E., Minneapolis, MN 55455}
\address[Ryu] {Department of Astronomy and Space Science, Chungnam National
University, Daejeon, 305-764, Korea}
\address[Park] {University of Minnesota Supercomputing Institute, 
1200 Washington Avenue S., Minneapolis, MN 55415}

\begin{abstract}
We have applied an effective numerical scheme for cosmic-ray transport to 
3D MHD simulations of jet flow in radio galaxies (see the companion paper
by Jones \etal ~1999). The marriage of 
relativistic particle and 3D magnetic field information allows us to construct
a rich set of ``synthetic observations'' of our simulated objects. The 
information is sufficient to calculate the 
``true'' synchrotron emissivity at a given frequency using explicit
information about the relativistic electrons. This enables us to produce 
synchrotron surface-brightness maps, including 
polarization. Inverse-Compton X-ray surface-brightness maps may also be 
produced. First results intended to explore the 
connection between jet dynamics and electron transport in radio lobes are
discussed.  We infer lobe magnetic field values by comparison of synthetically
observed X-ray and synchrotron fluxes, and find these ``inverse-Compton'' 
fields to be quite consistent with the actual RMS field averaged over the lobe.
The simplest minimum energy calculation from the synthetic observations also
seems to agree with the actual simulated source properties. 
\end{abstract}

\end{frontmatter}

\section{Introduction}
\label{intro}

Numerical simulations of extragalactic jets have been used for well over a
decade to shed light on the physical processes taking place within radio
galaxies.  There is an important 
and sometimes overlooked issue when making the comparison between numerical
simulations and observed sources: the link between dynamical processes
and the resulting emissions.  Synthetic observations 
can address this issue.  By synthetically observing a source
whose detailed physical structure is known beforehand, we can
gain insights into what real observations are actually telling us.

Attempts have been made to obtain emission characteristics from purely 
hydrodynamical simulations.  The earliest versions estimated
the local magnetic field strength by assuming it to be proportional to 
the density, providing a means to derive a crude
synchrotron ``pseudoemissivity.'' 
\citeasnoun{Smith85} made comparisons between observed hot spot morphologies 
and 
simulated hot spot dynamics in this way.
Yet purely hydrodynamical simulations can never truly model the jet 
synchrotron emission, as emphasized by \citeasnoun{Hughes96}, who found that
the distribution of hydrodynamic variables provides little insight into the 
observed intensity distribution in relativistic hydrodynamic simulations. 
Inclusion of even a passive vector magnetic field (e.g., Laing 1981; Clarke,
Norman, \& Burns 1989)
seemed to offer a big step forward, since the synchrotron emission reflects
in part the magnetic field distribution, and since it allows one to
calculate observable polarization properties. At this level synthetic
synchrotron brightness distributions show encouraging
morphological similarities to real radio galaxies.

However, synthetic radio observations from large-scale, 
fully 3-dimensional purely magnetohydrodynamical simulations 
still require critical arbitrary assumptions to model the emission.  
In addition, the emission spectrum cannot be obtained this way 
\cite{Clarke93}.  Only an explicit treatment
of the relativistic electron population can allow us to 
treat the emission properties of the simulation directly.  Jones, Ryu \& 
Engel (1999) (JRE99) and Jones, Tregillis \& Ryu (these proceedings) (JTR99)
describe a computationally
economical technique for relativistic particle transport within
multidimensional MHD flows.  This scheme can follow explicitly
diffusive acceleration at shocks and second-order Fermi acceleration in smooth
flows.  It also accounts for adiabatic losses, synchrotron
and inverse-Compton losses, and injection of fresh particles at shocks.

We have begun a series of 3-dimensional MHD simulations of
extragalactic radio jets utilizing this nonthermal particle transport scheme.
From the data produced by these radio galaxy simulations 
we can conduct a very rich set of synthetic observations, because for the first
time we are able to compute in each source volume element the number
of relativistic electrons at the energy radiating at a chosen frequency.
Presented here are some results from our first suite of synthetic observations,
which demonstrate the kinds of insights that can be obtained from this tool.
Related dynamical aspects of the simulations are described in JTR99.

\section{Synthetic Observation Methods}
\label{methods}

The combination of a vector magnetic field structure and a nonthermal particle
energy distribution makes it possible for us to compute an approximate 
synchrotron
emissivity, complete with spectral and polarization information, 
in every zone of the 
computational grid.  Surface-brightness maps for the optically-thin emission 
are then easily produced by ray tracing through the computational grid,
thereby projecting the source onto the plane of the sky.
We have also produced X-ray surface brightness maps in the same fashion, by
calculating the inverse-Compton (IC) emissivity from the interaction between 
the
cosmic microwave background radiation and our nonthermal electrons.  Future
work will include a more sophisticated treatment of the X-ray emission,
including synchrotron and AGN seed photons.

The synthetic observations can be imported into any 
standard image analysis package and analyzed like actual
observations. 
To make this exercise as realistic as possible we place the simulated object 
at an appropriate distance, currently 100 Mpc, and then convolve the computed surface
brightness distribution with typically sized Gaussian beams. We can then
explore a range of observable properties at multiple wavelengths,
multiple angular resolutions and so on.
The real power in this method comes from the fact that we can
then compare synthetically ``observed'' source properties with the actual
physical properties of the simulated objects.
We can, for instance, extract the true magnetic field strength and topology 
in a particular
region, particle and field ``filling factors'', or the distributions 
of particle spectral forms in an emitting volume.

\section{Discussion}
\label{Discussion}

Figure 1 illustrates ``observed'' radio synchrotron and X-ray IC images
constructed in this way from a 3-dimensional simulation described in JTR99.
Briefly, it represents emission from a 
light, supersonic jet, with density $10^{-2}$ 
of the ambient  medium, and internal Mach number of 8. For the other
parameters mentioned below this corresponds to a jet velocity of $0.1c$.  
The jet inflow slowly
precesses around a cone of opening angle $5^o$, to 
break cylindrical symmetry.  The incoming jet core radius was $r_j = 1$ kpc, and
the magnetic field on the jet axis was $B_0 = 1 \mu$G, with a gas pressure
there 100 times greater than the magnetic pressure. This leads to a
jet kinetic power of $\approx 10^{37}$ W. 
For this simulation a relativistic electron population with a power law
momentum index, $f(p) \propto p^{-4.4}$ (corresponding to a synchrotron
spectral index $\alpha = 0.7$) was brought onto the grid with
the jet flow. That population was subjected to acceleration at shocks, 
adiabatic and radiative losses. No additional nonthermal electrons
were injected at shocks, however.
In this simulation, the synchrotron cooling time for $\gamma = 10^4$ 
electrons radiating on the jet axis ($\nu \sim 300$ MHz)
is $3.7 \times 10^8$ years, whereas the analogous inverse-Compton 
($h\nu \sim 100$ keV) cooling time is $4.7 \times 10^7$ years. 
The synthetic observations presented here
correspond to a time when the jet has propagated approximately
$10^6$ years, rendering these cooling effects negligible.
The 1.4 GHz synchrotron luminosity $\nu L_\nu$ is 
$5.3 \times 10^{32}\times\delta_4$ W,
where $\delta_4 = 10^4\delta$, and $\delta$ is the ratio of nonthermal
to thermal electron densities at the jet orifice.

In what follows we have set $\delta_4 = 1$, motivated by 
estimates of injection efficiencies at nonrelativistic shocks 
\citeaffixed{TWJ99}{e.g.,}.
It gives the simulated synchrotron source a relatively low luminosity 
compared to typical FR2 objects.
It guarantees, however, that the relativistic electrons are 
passive.  In fact, the integrated pressure in relativistic electrons ($E > mc^2$) 
is everywhere less than 0.1\% of the thermal pressure.
Given that, the synthetic observations can all be scaled simply in terms of
$\delta$, or in some cases, are independent of $\delta$.
If, for comparison,
we assume from the same motivations that cosmic-ray ions
have $\sim 100$ times more energy than electrons, those nonthermal ions
could locally contribute up to $\sim 10$\% of the total pressure.
With these assumptions our simulated source is out of equipartition between
relativistic particles and magnetic fields, with the balance in favor of
particles.

The synchrotron synthetic observation in Figure 1 corresponds to a frequency
of 1.4 GHz, and the X-ray image to an energy of 10 keV. 
With the source
distance assumed to be 100 Mpc
the synchrotron flux over the source is about $330\times\delta_4$ mJy.
The 10 keV inverse-Compton flux is $2.4\times10^{-10}\times\delta_4$ Jy
corresponding to 
a 10 keV luminosity $\nu L_\nu$ of $6.5\times10^{32}\times\delta_4$ W.

\begin{figure}
\begin{center}
\includegraphics*[width=14cm]{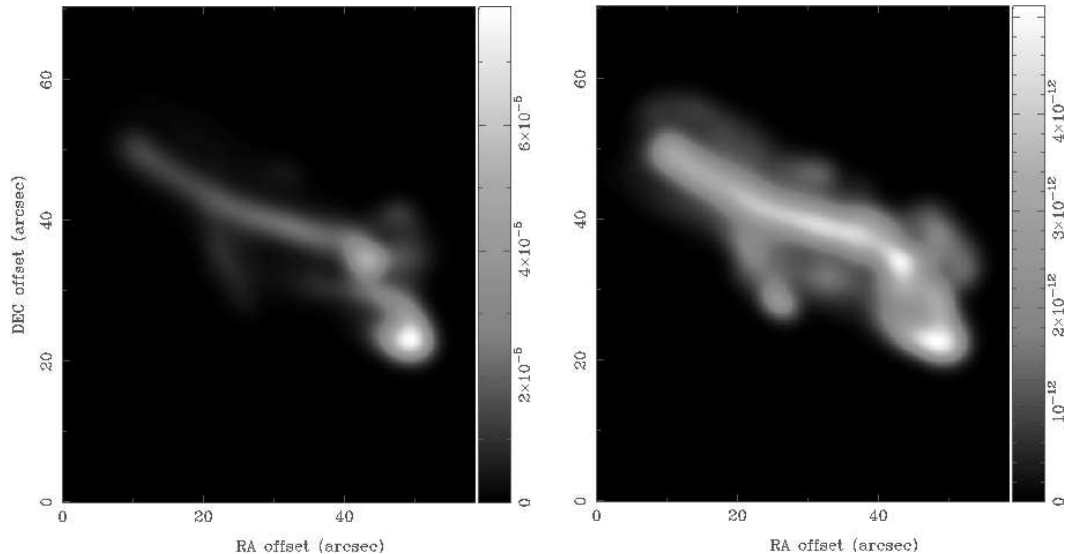}
\end{center}
\caption{Synthetic surface brightness maps constructed as
described in the text and analyzed with the MIRIAD package.  The left panel
is 1.4GHz synchrotron emission; the right panel is 10keV inverse-Compton 
X-ray emission.  The linearly-scaled colorbar values represent Janskys per
3.5'' beam.}
\label{fig_IXsynth}
\end{figure}

In the images one can clearly separate the jet, a terminal hot spot, 
a secondary hot spot and the diffuse lobe.
The contrast in the radio image between the jet core and the outer lobe 
is higher than sometimes observed in real sources.  
There are two primary reasons
why the jet is so prominent. First, the entire cosmic-ray electron population
passes down the jet in this simulation, so the density of relativistic
electrons is higher there than anyplace else. (See the 2-dimensional
simulations of JRE99 to measure the importance of this
modeling choice.) Since the magnetic field is stronger inside the jet
than in much of the lobe volume, the synchrotron emissivity is relatively
large there. In addition, the source in Figure 1 is quite young dynamically,
so that a line of sight path length through the jet is roughly 20\% of the
path through the lobe. As this source aged that comparison would
change significantly in favor of the lobe path length, of course.

While there is an overall correspondence between the radio and X-ray images, 
there are some significant differences.  
The jet/lobe and hot-spot/lobe contrasts are much more 
pronounced in the synchrotron map.
Calculating the average surface brightness in the jets and in the lobes
of the two maps, we find that this ratio is about three times
higher in the radio map than in the X-ray map.  
While the radio hot spots do appear to have X-ray bright counterparts, there
are also bright X-ray regions without corresponding radio enhancements.
The inverse-Compton emission arises
from a particle population of fixed energy ($\sim 1$ GeV at 10 keV), 
while the synchrotron emission at fixed frequency does not.  
Thus, the surface brightness of IC emission from the CMB simply reflects the 
column density of electrons in a fixed energy range, while the 
synchrotron is biased to regions where the fields are strongest, both
because radiative power is greater there, but also because one sees
emission from relatively more plentiful lower energy electrons.
In cases where radiative losses are significant in the GeV range so 
that the spectra steepen, these differences would become even more dramatic.

Comparison of the two prominent hot spots reveals some interesting contrasts.
The larger primary hot spot at the right edge of the radio lobe corresponds
to an extremely complicated set of structures near the jet terminus.
While it is nominally associated with the end of the jet
flow, the jet terminal shock is very small and influences the
spectrum of only a tiny portion of the electrons in the hot spot.
In fact, much of the emergent jet flow has not passed through any
strong shock before reaching the cocoon.
Some of the electron population shows signs of significant shock
acceleration, but the pattern is not simple 
(see also JTR99.)
Over all, the influence of shock acceleration on the brightness and
spectrum of the primary hot spot is quite small.
The enhanced emission is mostly due to magnetic field 
amplification in the complex flows there rather than particle acceleration.

On the other hand, there is clear evidence for particle acceleration in 
the secondary hot spot further back in the lobe,  which has a much 
different character.  This hot spot has a distinctly 
flatter spectrum 
($\alpha \approx 0.6$) than the jet ($\alpha = 0.7$).
The surface brightness map would make it appear that the jet flows 
directly into this hot spot.  However, this is an accident of
projection. The high emissivity volume really is well outside the jet,
and examination of the plasma streak lines reveals that this hot spot
actually corresponds to flow downstream of a moderately strong shock
formed in the cocoon ``backflow'', as discussed in JTR99.

One of the greatest potential applications of synthetic observations is 
comparisons between the real physical properties of a simulated object
and observationally inferred source properties, which are generally based on
convenient, simplified assumptions.
Here we provide a couple of preliminary examples, from the simulation
described above and in JTR99.
From comparison of the integrated radio synchrotron and X-ray IC
measured fluxes,
we can infer a magnetic field  within the source \citeaffixed{Harris79}{e.g.,}.
For our simulated source and synthesized observations the resulting
``inverse-Compton'' field is\\
\centerline{$B_{ic} =  1.0~\mu {\rm G}$,}\\
which matches very well the actual volume-averaged RMS field within the lobes;
namely, $B_{rms} = 1.2~\mu$G. That is a very encouraging result,
especially since the magnetic field in the source is really
quite filamentary, with an ``intermittency'' 
($\left < B^4\right >/(\left < B^2\right >^2) \approx 5.7$ at the time shown. 
The maximum field value in the lobes is $12~\mu$G and the minimum
field value is $0.003~\mu$G. 

We also can apply standard  minimum-energy arguments (between
nonthermal particles and magnetic fields) to the observed
synchrotron flux for an associated magnetic field
strength.  Again integrating over the source, this gives
\citeaffixed{Feigelson95}{e.g.,}\\
\centerline{      $B_{me} =	(5.4~\mu {\rm G}) 
\times ((1+k)/(101\eta))^{2/7} \times \delta_4^{2/7}$,}\\
where $\eta$ is the filling factor and $k$ is the ratio of nonthermal
ion to electron energies.
Even with $\eta = 1$ this minimum energy magnetic field value exceeds 
the actual RMS field and the IC estimate by roughly a factor of 5, warning us
that the field is not in a minimum-energy configuration.
More realistic
choices for the filling factor would increase that difference.
This result is very consistent with the actual simulated source properties,
because, as noted earlier, with $\delta_4 = 1$ and $k = 100$,
nonthermal particle energies do actually exceed the global magnetic
field energy by an order of magnitude or more. Thus, again, we find
this an optimistic result, because it suggests the combined use of
IC and minimum energy magnetic field estimates may be able to
give us reliable information about the actual fields and the 
actual partitioning of energy between particles and fields.

This work was supported at the University of Minnesota by the NSF through
grant AST96-16964 and by the University of Minnesota Supercomputing
Institute. DR was supported in part by KOSEF through grant 981-0203-011-02.

\end{document}